\documentclass[twocolumn,amsmath,amssymb,prl]{revtex4}
\usepackage{color}
\usepackage[pdftex]{graphicx}
\usepackage{dsfont}

\newcommand{\rr}{\mathbf{r}}

\newcommand{\rrho}{\boldsymbol{\rho}}
\newcommand{\kk}{\mathbf{k}}
\newcommand{\qq}{\mathbf{q}}

\input epsf
\begin{document}
\title {Half-Quantum Vortex Molecules in a Binary Dipolar Bose Gas}
\author{Wilbur E. Shirley$^{1,2}$, Brandon M. Anderson$^{2}$, Charles W. Clark$^{2}$ and Ryan M. Wilson$^{2}$}
\affiliation{$^1$Department of Physics, University of Illinois at Urbana-Champaign, Urbana, Illinois 61801, USA}
\affiliation{$^2$Joint Quantum Institute, National Institute of Standards and Technology and the University of Maryland, College Park, Maryland 20742, USA}
\date{\today}
\begin{abstract}
We study the ground state phases of a rotating two-component, or binary Bose-Einstein condensate, wherein one component possesses a large magnetic dipole moment.  % The elementary topological excitations in this system are half-quantum vortices (HQVs), which may be nucleated by mechanical rotation or an appropriately engineered synthetic gauge field.  
A variety of non-trivial phases emerge in this system, including a half-quantum vortex (HQV) chain phase and a HQV molecule phase, where HQVs of opposite charge bind at short distances.  We attribute the emergence of these phases to the development of a minimum in the adiabatic HQV interaction potential, which we calculate explicitly.  We thus show that the presence of dipolar interactions in this system leads to a rich phase diagram, and the formation of HQV molecules.
% This system supports self-assembled structures of topological excitations, or half-quantum vortices when subject to  rotation or an appropriately engineered gauge field.  We explore the phase diagram of this 
% We consider a rotating two-component Bose-Einstein condensate in quasi-two dimensional geometry, wherein one component exhibits dipole-dipole interactions. We model numerically the interaction potential between a half-quantized vortex (HQV) in the dipolar species and a HQV in the other species. We find that for sufficiently strong dipolar interactions a bound state between HQVs occurs as a result of rotonic features induced by the dipolar vortex. We then simulate a rapidly rotating system confined in an oblate harmonic trap and observe novel ground state vortex configurations including HQV molecules and chains of bound vortices. Finally we present a phase diagram which elucidates the effects of dipolar interactions on the planar vortex geometry.
\end{abstract}
\maketitle

Topological defects are objects of fundamental importance in modern physical theories. %, ranging from cosmology to condensed matter.
 % They represent field configurations that are homotopically distinct from the underlying vacuum
In superfluids, they take the form of quantized vortices, which are line-defects around which the phase of the order parameter winds by an integer multiple of $2\pi$~\cite{Gross61,Fetter67,Fetter01}.  Vortices can manifest in the ground state of a superfluid under rotation~\cite{Yarmchuk79,Madison00,AboShaeer01,Haljan01,Zhang05,Komineas07,Cooper2008,Fetter09,Malet11} or in the presence of a magnetic gauge field~\cite{Abrikosov57,Trauble67,Blatter94,Lin2009,Dalibard2011,Goldman13arXiv}; they play a fundamental role in the turbulent dynamics and thermalization of far-from-equilibrium states~\cite{Schwarz88,Kozik04,Nowak12,Bradley12}, and they mediate the transition from a superfluid to a normal fluid in two-dimensions (2D)--the Berezinsky-Kosterlitz-Thouless (BKT) transition~\cite{Berezinskii70,Berezinskii71,Kosterlitz73,Hadzibabic06,Posazhennikova06}. % These phenomena all rely crucially on the nature of vortex interactions in the superfluid state.

In recent years, experimental progress in the laser cooling and trapping of neutral atoms has permitted the experimental investigation of superfluidity and vortices in highly versatile, controllable environments~\cite{Dalfovo99}.  For example, the formation of an Abrikosov, or triangular vortex lattice was demonstrated in a rotating atomic Bose-Einstein condensate (BEC)~\cite{AboShaeer01}, and the vortex-antivortex unbinding mechanism was shown to be responsible for the BKT transition in a highly oblate Bose gas~\cite{Hadzibabic06}.
%  These systems allow us to explore more exotic and unconventional models, as well.
Today, cold atoms provide us with tools to % transcend the regime of traditional condensed matter systems.  
explore less conventional superfluid systems, including BECs of atoms with large permanent magnetic dipole moments such as Cr, Er, and Dy~\cite{Griesmaier05,Pasquiou11,Lu11,Aikawa12,Baranov12}, % which are predicted to exhibit a roton-like quasiparticle excitation~\cite{}, 
and two-component (binary) BECs, which are host to coreless,  or ``half-quantum'' vortex (HQV) topological defects~\cite{Hall98,Schweikhard04,Kasamatsu05rev,Papp08,Mason11}.  Due to the distinctive character of their interactions, 
% HQVs were shown to crystalize into a square lattice when the binary BEC is rapidly rotated near its immiscible-miscible transition (IMT) threshold~\cite{}.  Additionally, 
HQVs are predicted to mediate a BKT transition in the ``spin'' order of the binary Bose gas~\cite{Mukerjee06,Shi11}, and to form bound pairs when the two components are coherently coupled by a resonant optical field~\cite{Kasamatsu04,Kasamatsu05}.

\begin{figure}[b]
\includegraphics[width=.85\columnwidth]{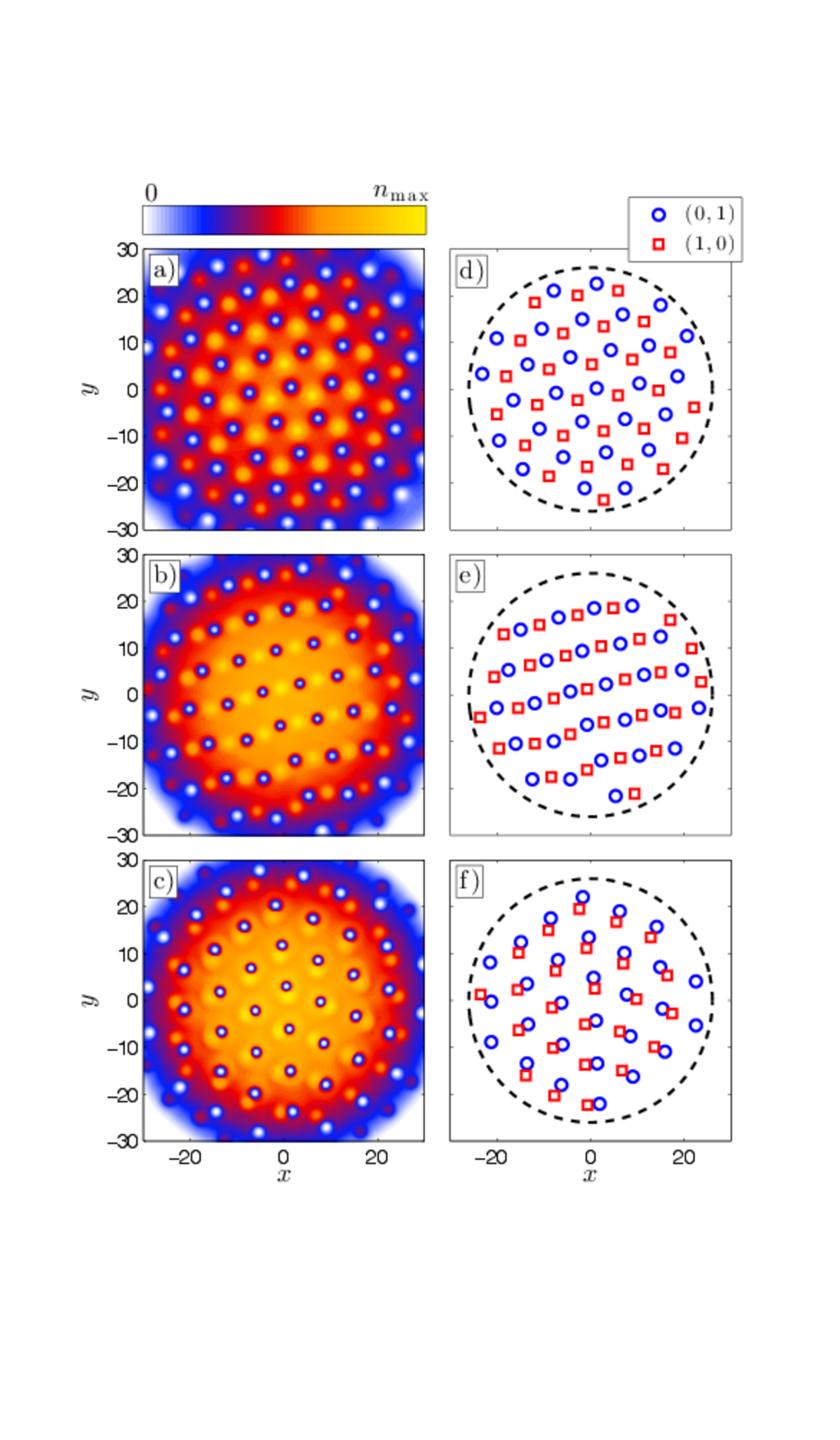}
\caption{\label{fig:den} (color online).  Emergent HQV phases of the rotating binary dipolar BEC for  $N \tilde{g} = 10^4$,  $\Delta = 0.3$, $\Omega = 0.5 \omega_\rho$, and $\lambda=10$.
(a)-(c) Densities of non-dipolar component, $n_2(\rrho)$, and (d)-(f) locations of $(1,0)$ (red squares) and $(0,1)$ (blue circles) HQVs for (a),(d) $\tilde{g}_d / \tilde{g} = 0$, (b),(e) $\tilde{g}_d / \tilde{g} = 0.65$, and (c),(f) $\tilde{g}_d / \tilde{g} = 1.25$.  For clarity, we only show HQVs within a radius of $25 l_z$ from the origin, marked by the black dashed line.}
\end{figure}

In this Letter, we consider a binary BEC of highly polarized dipolar atoms in an oblate geometry, in which the single-component dipolar BEC is predicted to exhibit a roton-like minimum in its quasiparticle dispersion~\cite{ODell03,Santos03}.  We explore the phase diagram of this system in the presence of rotation~\cite{Zhao13,Ghazanfari14}, and identify a number of exotic phases that emerge when one component is non-dipolar, including a HQV chain phase and a HQV molecule phase (shown in Fig.~\ref{fig:den}), where HQVs bind at a short distance in the \emph{absence} of coherent coupling.  We calculate the inter-HQV potential explicitly, and find that it exhibits a short-range minimum when the dipole-dipole interactions (DDIs) are sufficiently strong, which we attribute to the presence of a spin-wave roton feature in the quasiparticle spectrum of the gas.  Thus, we demonstrate that the binary dipolar BEC exhibits HQV molecules, which emerge in the rich ground state phase diagram of the system, and should play a critical role in its dynamical and superfluid properties.

\emph{Formalism}--  The binary Bose gas is described by the many-body Hamiltonian
\begin{align}
\label{H}
\hat{\mathcal{H}} = \hat{\mathcal{H}}_0 + \hat{\mathcal{H}}_\mathrm{int},
\end{align}
where $\hat{\mathcal{H}}_0 = \sum_{\alpha=1,2} \int d\rr \hat{\Psi}^\dagger_\alpha(\rr) \hat{H}_\alpha (\rr) \hat{\Psi}_\alpha(\rr)$, $\hat{\Psi}_\alpha(\rr)$ ($\hat{\Psi}^\dagger_\alpha(\rr)$) is the Bose field annihilation (creation) operator for atoms in component $\alpha$, and $\hat{H}_\alpha(\rr)$ is the single particle Hamiltonian
\begin{align}
\label{H1}
\hat{H}_\alpha(\rr) = - \frac{\hbar^2}{2m_\alpha}\nabla^2 + \frac{1}{2} m_\alpha \omega_z^2 \left( z^2 + \rho^2 / \lambda^2 \right) - \Omega_\alpha L_z,
\end{align}
where $m_\alpha$ is the atomic mass, $\lambda = \omega_z / \omega_\rho$ is the trap aspect ratio, $\omega_{z}$ ($\omega_\rho$) is the axial (radial) trap frequency, $L_z =  -i \hbar ( x \partial_y - y \partial_x)$, and $\Omega_\alpha$ is the effective rotation frequency of component $\alpha$.  The last term in Eq.~(\ref{H1}) could result from a mechanical rotation of the system (in which case $ \Omega_1 = \Omega_2$, and the Hamiltonian is expressed in the rotating frame) or a Raman laser-induced synthetic gauge field applied to component $\alpha$~\cite{Lin2009,Dalibard2011,Goldman13arXiv}.  The second term in Eq.~(\ref{H}) is the interaction Hamiltonian~\cite{Goral02},
\begin{align}
\label{Hint}
\hat{\mathcal{H}}_\mathrm{int} =  \frac{1}{2} \sum_{\alpha, \beta = 1,2} \int d\rr \int d\rr^\prime\hat{n}_\beta(\rr^\prime) V_{\alpha \beta}(\rr-\rr^\prime) \hat{n}_\alpha(\rr),
\end{align}
where $\hat{n}_\alpha(\rr) = \hat{\Psi}^\dagger_\alpha(\rr) \hat{\Psi}_\alpha(\rr)$ and $V_{\alpha \beta} (\rr) $ is the two-body interaction potential.  For highly magnetic atoms polarized along the $z$-axis by an external field, this potential can be written in terms of an isotropic short-range pseudo-potential and a long-range DDI, 
\begin{align}
V_{\alpha \beta}(\rr) = g_{\alpha \beta} \delta(\rr) + d_\alpha d_\beta \frac{1 - 3 \cos^2 \theta_\rr}{|\rr|^3},
\end{align}
where $\theta_\rr = \mathbf{e}_z \cdot \rr / |\rr|$, $d_\alpha$ is the magnetic dipole moment of atoms in component $\alpha$, $g_{\alpha \beta} = 2\pi \hbar^2 a_{\alpha \beta} / \tilde{m}_{\alpha \beta} $, $a_{\alpha \beta}$ is the $s$-wave scattering length for collisions between atoms in components $\alpha$ and $\beta$, and $\tilde{m}_{\alpha \beta} = m_\alpha m_\beta / (m_\alpha + m_\beta)$ is a reduced mass.  Here, we have ${g}_{12} = {g}_{21}$.  

When $\lambda \gg 1$, corresponding to a highly oblate geometry, %and $\hbar \omega_z$ is large compared to other relevant energy scales in the system, %and $\hbar \omega_z$ is large compared to other energy scales in the system, 
we can construct an effective low-energy, quasi-two-dimensional (Q2D) theory by separating a common transverse mode from the field operators, $\hat{\Psi}_\alpha(\rr) = \hat{\psi}_\alpha(\rrho) f_\alpha(z)$, and integrating the $z$-coordinate out of the Hamiltonian.  We take $f_\alpha(z) = \exp \left[ - z^2 / 2 l_\alpha^2 \right ] / \pi^\frac{1}{4} \sqrt{l_\alpha}$ where $l_\alpha = \sqrt{\hbar / m_\alpha \omega_z}$, corresponding to the single particle ground state of a 1D harmonic oscillator.  % While this approximation is only quantitatively exact in the limit of vanishingly weak interactions, it captures the qualitative features of interacting gases when axial dynamics can be neglected, as is the case here.  
Further, in the ultracold, dilute limit, it is natural to introduce the mean-field condensate order parameters $\langle \hat{\psi}_\alpha(\rrho) \rangle \simeq \phi_\alpha(\rrho)$, where $\phi_\alpha(\rrho)$ is normalized to the condensate occupation, $\int d\rrho |\phi_\alpha (\rrho)|^2 =N_\alpha$.  In proceeding, we specialize to the ``balanced'' case with $m_1 = m_2 = m$, $\Omega_1 = \Omega_2 = \Omega$, $N_1 = N_2 = N$, and we work in dimensionless units by scaling with the trap energy $\hbar \omega_z$ and characteristic oscillator length $l_z = \sqrt{\hbar / m \omega_z}$ where appropriate.  In the mean-field approximation, the energy of the Q2D binary BEC, up to an overall constant, is given by 
\begin{align}
\label{Q2DE}
E &= \sum_{\alpha = 1,2} \int d\rrho \, \phi^*_\alpha(\rrho) \left\{ -\frac{ \nabla_\perp^2}{2}  + \frac{\rho^2}{2\lambda^2}  - \Omega L_z \right\} \phi_\alpha(\rrho) \nonumber \\
&+ \frac{1}{2}\sum_{\alpha,\beta = 1,2}\int d\rrho \, n_\alpha(\rrho) \left\{ \tilde{g}_{\alpha \beta} n_\beta(\rrho) + \tilde{g}_{d,\alpha \beta}  \Phi_{\alpha \beta}(\rrho) \right\},
\end{align}
where $\nabla_\perp^2 = \partial_x^2 + \partial_y^2$, $n_\alpha (\rrho) = |\phi_\alpha(\rrho)|^2$, $\tilde{g}_{\alpha \beta} = g_{\alpha \beta} / \sqrt{2\pi}$, $\tilde{g}_{d,\alpha \beta} = \sqrt{8\pi} d_\alpha d_\beta / 3 $, and $ \tilde{g}_{d,\alpha \beta} \Phi_{\alpha \beta} (\rrho)$ is a mean-field potential generated by the dipoles of component $\beta$.  We calculate $\Phi_{\alpha \beta}(\rrho)$ by employing the convolution theorem, $\Phi_{\alpha \beta}(\rrho) = \mathcal{F}^{-1}_{2D} \left[ \tilde{n}_\beta(\kk) F(\kk / \sqrt{2})  \right]$, where $\tilde{n}_\alpha(\kk) = \mathcal{F}_{2D}[ n_\alpha(\rrho)]$, $F(\qq) = 2 - 3 \sqrt{\pi} q e^{q^2} \mathrm{erfc}(q)$, $\mathcal{F}_\mathrm{2D}$ is the 2D Fourier transform operator, and $\mathrm{erfc}$ is the complementary error function~\cite{Fischer06,Nath09,Ticknor11,Wilson12}.

The energy functional~(\ref{Q2DE}) is invariant under the global gauge transformation $\phi_\alpha(\rrho) \rightarrow \phi_\alpha(\rrho) e^{i\varphi}$, reflecting the $U(1)_1 \otimes U(1)_2$ symmetry of the Hamiltonian~(\ref{H}).  The emergence of the condensate order parameters $\phi_\alpha$ (with well-defined global phases) breaks this symmetry.  The first homotopy group of the order parameter space is thus $ \mathbb{Z} \oplus \mathbb{Z}$, so there exist two winding numbers that classify the topological line defects in the binary BEC, corresponding to vortex excitations in the two components~\cite{Eto11}.  We denote a vortex with winding number $l$ in component 1 (2) as $(l,0)$ ($(0,l)$).  If the order parameters are represented locally in the spinor notation $e^{i \varphi_m} ( |\phi_1| e^{i \varphi_s} , |\phi_2| e^{-i \varphi_s} )^\mathrm{T}$, then the $2\pi$ phase windings of the singly quantized  $(1,0)$ and $(0,1)$ vortices correspond to a $\pi$ winding of the phase $\varphi_m$.  We thus refer to these topological excitations as ``half-quantum'' vortices (HQVs).  

% In proceeding to study the properties of this system, which involves a large parameter space, we specialize to the case with $\tilde{g} = \tilde{g}_{11} = \tilde{g}_{22}$, and $d_2$=0, so component 2 is non-dipolar.  We define $\tilde{g}_d = \tilde{g}_{d,11}$ to be the DDI coupling for component 1.  %Additionally, we introduce the parameter $\Delta = 1 - \tilde{g}_{12}^2 / \tilde{g}^2$ to characterize the relative contact interaction strengths.  % In the non-dipolar, non-rotating, homogeneous case (with $\lambda \rightarrow \infty$), $\Delta > 0$ corresponds to a miscible ground state and $\Delta<0$ corresponds to an immiscible ground state, where the components prefer to spatially separate.    
% In the remainder of this Letter, we work in dimensionless units by scaling with the trap energy $\hbar \omega_z$ and characteristic oscillator length $l_z = \sqrt{\hbar / m \omega_z}$ where appropriate.  

% In previous work~\cite{Wilson12}, this system was shown to exhibit a longitudinal spin-wave rotonization in its spectrum of elementary quasiparticle excitations (if an effective spin degree of freedom is defined as $S_z(\rrho) = n_1(\rrho) - n_2(\rrho)$) in the absence of rotation ($\Omega=0$) and a ``roton'' immiscible-miscible phase transition (IMT), where the components spatially separate at a length scale $\sim a_z$, in contrast to the long-wavelength ``phonon'' IMT in non-dipolar binary BECs.  

\begin{figure}[t]
\includegraphics[width=.95\columnwidth]{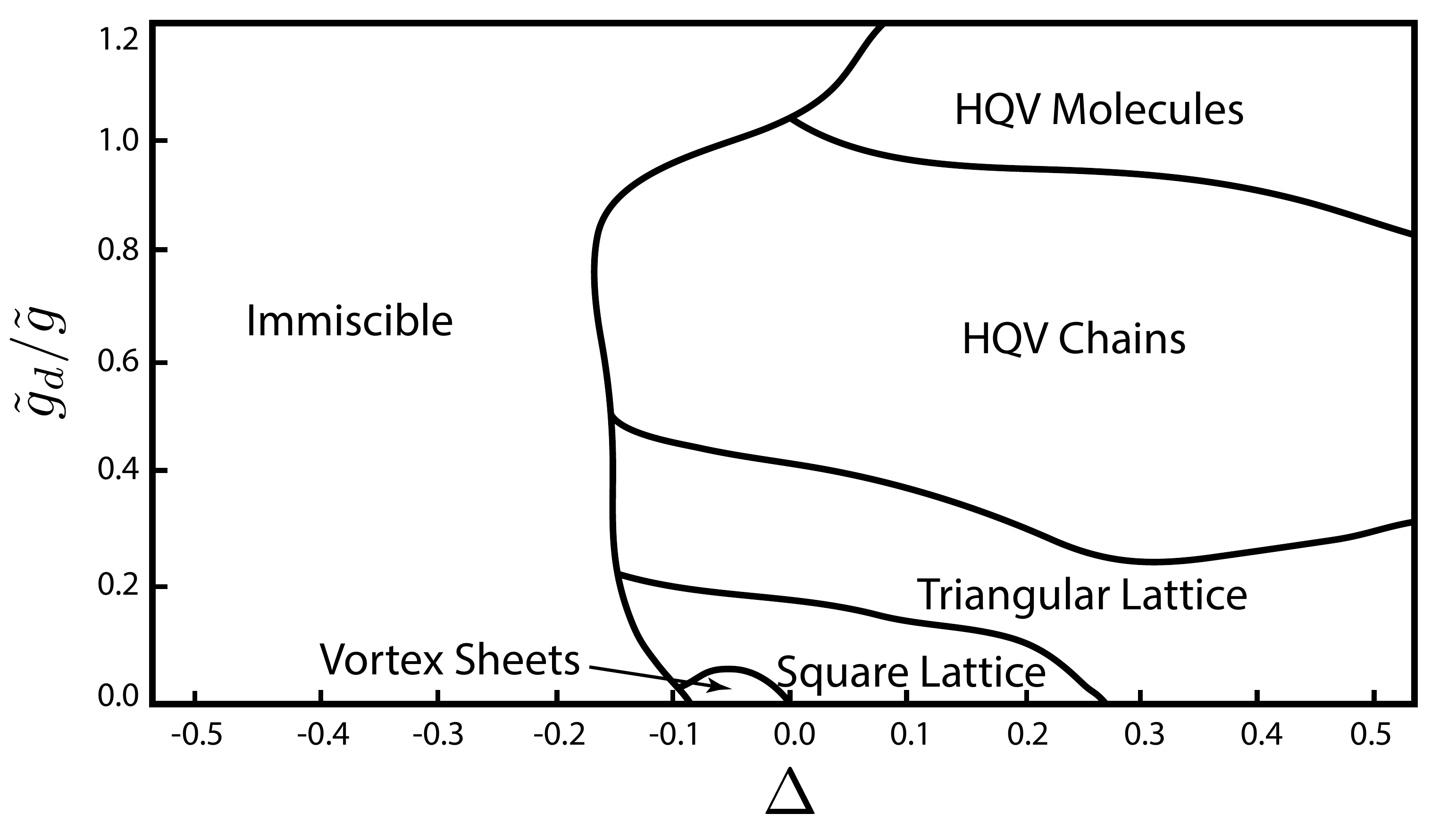}
\caption{\label{fig:pd} (color online).  Phase diagram of the ``balanced'' binary BEC with a trap aspect ratio $\lambda = 10$, a rotation frequency $\Omega = 0.5 \omega_\rho$, and $N \tilde{g} = 10^4$. See text for discussion of phases.    }
\end{figure}

\emph{Phase Diagram}--  To study the Bose condensed phase, we numerically minimize the energy functional~(\ref{Q2DE}) % by representing the condensate wave functions on a discrete Fourier grid, 
% using imaginary time propagation~\cite{} 
and find the stationary, dynamically stable ground state configurations.  We specialize to the case where $ \tilde{g}_{11} = \tilde{g}_{22} = \tilde{g}$ and component 2 is non-dipolar, so $d_2=0$.  We define $\tilde{g}_d = \tilde{g}_{d,11}$ to be the DDI coupling for component 1.
In Fig.~\ref{fig:pd}, we show a phase diagram for this system as a function of $\Delta = 1 - \tilde{g}_{12}^2 / \tilde{g}^2$ and $\tilde{g}_d / \tilde{g}$ for $N \tilde{g} = 10^4$, a trap aspect ratio $\lambda=10$, and a rotation frequency $\Omega = 0.5 \omega_\rho$, which is well above the critical frequency for HQV nucleation.

For a range of $\Delta < \Delta_\mathrm{crit}$, the strength of the repulsive interspecies contact interaction overwhelms the intraspecies interactions and the components spatially separate, forming an immiscible phase.  For small $\tilde{g}_d$, the onset of immiscibility occurs at the largest length scale available in the system, being the size of the condensate.  For larger $\tilde{g}_d$, however, the transition to immiscibility is qualitatively different, occurring on a much smaller length scale $\sim l_z$, and leading to the formation of density bubbles and stripe phases.  Ref.~\cite{Wilson12} showed that this transition can be attributed to the softening of a roton-like quasiparticle at wave number $\sim l_z^{-1}$, which drives fluctuations of an effective longitudinal spin $S_z (\rrho) = n_1 (\rrho) - n_2 (\rrho) $.   In the following, we discuss how this spin-wave roton plays a profound role not only near the immiscible-miscible transition threshold, but deep in the miscible regime.  For details regarding the immiscible phases of rotating binary BECs, see Refs.~\cite{Kasamatsu05rev,Mason11} and references therein.

% We label this mode in Fig. (), where we show the longitudinal branch for $n_0 \tilde{g} = n_0 \tilde{g}_d = 2$ and $n_0 \tilde{g}_{12} = 1.8$

For $\Delta > \Delta_\mathrm{crit}$, the two components are miscible, and exhibit a rich variety of HQV configurations.  The phases at small $\tilde{g}_d$ are familiar from previous studies of rotating binary condensates.  In the ``vortex sheet'' phase, $(1,0)$ HQVs align in a winding sheet, which is interwoven with a sheet of $(0,1)$ HQVs~\cite{Kasamatsu09}.  In the square and triangular lattice phases, the HQVs form lattice structures with rhombic and triangular unit cells, respectively, and are staggered such that the $(1,0)$ HQVs are located within the unit cells of the $(0,1)$ HQV lattice, and vice versa~\cite{Mueller02}.  We show the density $n_2(\rrho)$ and the locations of the $(1,0)$ and $(0,1)$ HQVs in Figs.~\ref{fig:den}(a) and~\ref{fig:den}(d), respectively, for a binary BEC with $\tilde{g}_d = 0$ and $\Delta = 0.3$, corresponding to the triangular lattice phase.  The locations of the $(0,1)$ HQVs can be identified by the points where $n_2(\rrho)$ vanishes.  In the miscible regime with $\tilde{g}_{12}>0$, it is energetically favorable for the atoms in component 2 to occupy the $(1,0)$ cores, so the locations of the $(1,0)$ HQVs can be identified by the local maxima of $n_2(\rrho)$.  % This is in contrast to the case with $\tilde{g}_{12} < 0$, in which the $(1,0)$ and $(0,1)$ HQVs prefer to occupy the same point in space~\cite{Mueller02}  
In practice, we locate the $(1,0)$ and $(0,1)$ HQVs by finding the points around which the phases of $\phi_1(\rrho)$ and $\phi_2(\rrho)$ wind by $2\pi$, respectively.

As $\tilde{g}_d$ is increased, the HQV lattice configurations undergo a striking change.  For an intermediate DDI strength $\tilde{g}_d / \tilde{g} = 0.65$ (with $\Delta = 0.3$), as shown in Figs.~\ref{fig:den}(b) and~\ref{fig:den}(e), the $(1,0)$ and $(0,1)$ HQV lattices shift, and their unit cell boundaries align to form stripes of HQVs, or HQV chains.  % The emergence of this phase suggests that it is not energetically favorable for the $(0,1)$ and $(1,0)$ HQVs to be maximally separated, as they are in the square and triangular lattice phases.  
The HQV chain phase occupies a large region in the phase diagram, as seen in Fig.~\ref{fig:pd}.  As the DDI strength is further increased to $\tilde{g}_d / \tilde{g} = 1.25$, as shown in Figs.~\ref{fig:den}(c) and~\ref{fig:den}(f), adjacent $(1,0)$ and $(0,1)$ HQVs form pairs, indicating that these HQVs experience a mutual \emph{attraction} at a length scale on the order of the HQV pair size.
% suggesting that the effective interaction potential between these HQVs, $U_{(1,1)}(\rrho)$, has a local minimum at a length scale corresponding to the size of a HQV pair.  
% As $\tilde{g}_d $ is continually increased, the system undergoes a transition to a roton immiscible phase, and eventually collapses when $\tilde{g}_d \gtrsim 2 \tilde{g} $ due to the attractive part of the DDI overwhelming the repulsive contact interaction.  This collapse is akin to the roton collapse in scalar dipolar condensates in oblate geometries~\cite{}. 
% To draw the phase boundaries in Fig.~\ref{fig:pd}, we analyzed the qualitative structure of the system near the center of the cloud.

\emph{HQV Interactions}--  To understand why the HQV chains and pairs emerge, we study the interactions between $(1,0)$ and $(0,1)$ HQVs, which together form a $(1,1)$ system. % In the case where these vortices are stationary and separated by a distance $R$, we can use an adiabatic potential, $U_{(1,1)}(R)$, to characterize their interactions. 
Specifically, we calculate an adiabatic potential for the $(1,1)$ system in a homogeneous  geometry ($\lambda = \infty$) % with an integrated 2D density per component $n_0$ 
by numerically minimizing the functional~(\ref{Q2DE}) under the constraint that the HQVs are ``pinned'' at points $\rrho_{(1,0)}$ and $\rrho_{(0,1)}$, which are separated by a distance $R$.  In practice, we set $\Omega= 0$ and initialize our minimization procedure with states $\phi^{(0)}_1$ and $\phi^{(0)}_2$ that have $2\pi$ phase winding and vanishing density at the points $\rrho_{(1,0)}$ and $\rrho_{(0,1)}$, respectively. %, where they vanish, $|\phi^{(0)}_1(\rrho_{(1,0)}| = |\phi^{(0)}_2 (\rrho_{(0,1)})| = 0$.  
We pin the HQVs with strong,  localized state-dependent potentials, so they remain stationary throughout the energy minimization.  On a  grid of points $\rrho_i$, we use the pinning potential $V_0 \delta_{\rrho_i, \rrho_{(1,0)}}$ ($V_0 \delta_{\rrho_i, \rrho_{(0,1)}}$) for the $(1,0)$ ($(0,1)$) HQV.  Our results are independent of $V_0$ as long as it is sufficiently larger than other energy scales in the system.  % Once an energy-minimizing configuration is achieved, 
The adiabatic potential is given by
\begin{align}
\label{U}
U_{(1,1)}(R)  = E_{(1,1)}(R) - E_{(1,0)} - E_{(0,1)} + E_0,
\end{align}
where $E_{(1,1)}(R)$ is the total energy of the system with a $(1,1)$ pair separated by a distance $R$, $E_{(1,0)}$ ($E_{(0,1)}$) is the energy of a single HQV of type $(1,0)$ ($(0,1)$), and $E_0$ is the energy with no HQVs present~\cite{Eto11,Mulkerin13}.   
In the limit $\tilde{g}_d = 0$, this method agrees with the analytic result $U_{(1,1)}(R) = \frac{\pi}{4} \tilde{g}_{12} \mathrm{ln} \frac{R}{\xi} / ( n_0 \tilde{g}^2 \Delta R^2 )$ when $R \gg \xi$, where $\xi$ is a short-range cutoff~\cite{Eto11} and $n_0$ is the integrated 2D density of each component.

\begin{figure}[t]
\includegraphics[width=.85\columnwidth]{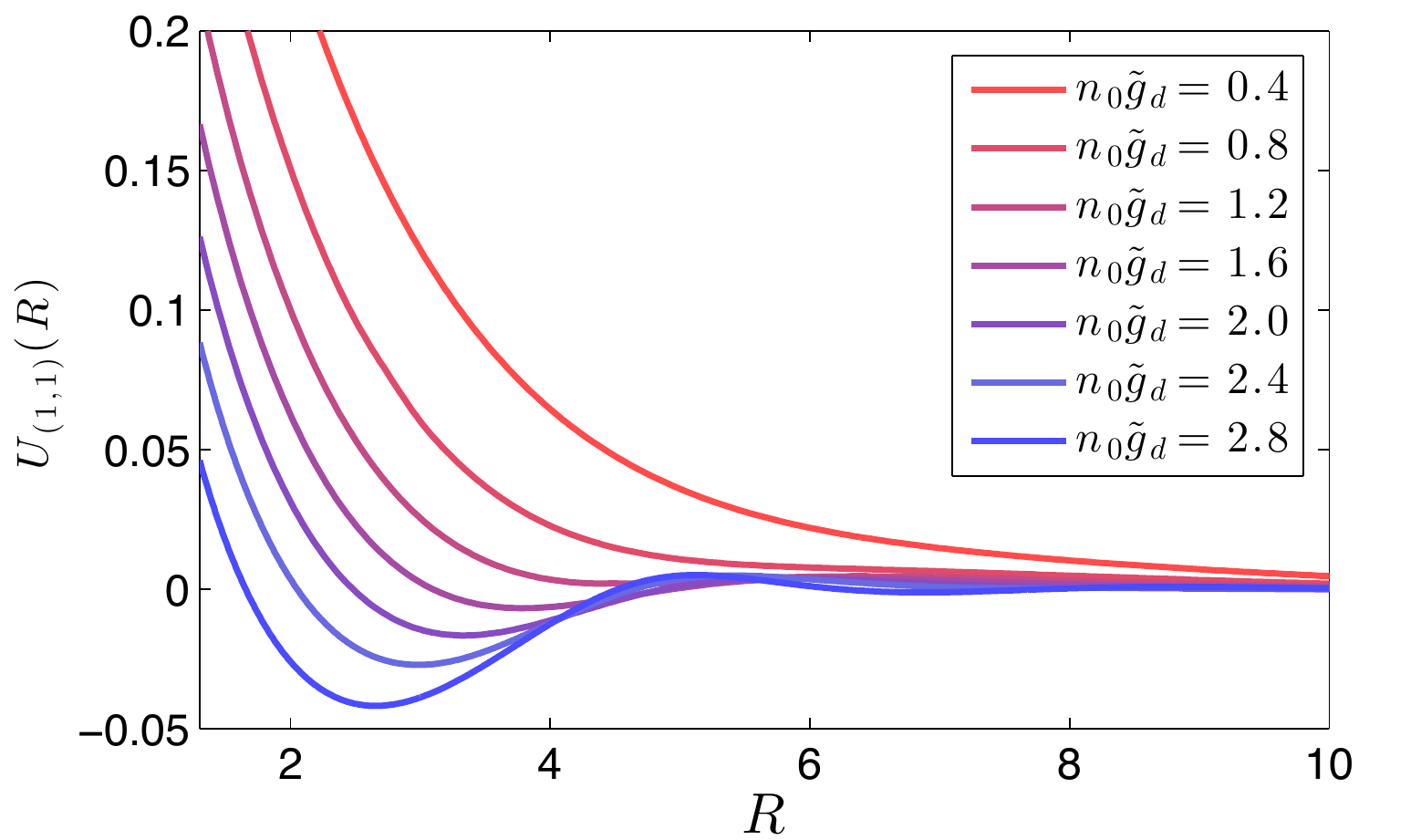}
\caption{\label{fig:hqvpot} (color online).  Adiabatic interaction potentials $U_{(1,1)}(R)$ (as defined in Eq.~(\ref{U})) of a $(1,1)$ HQV system for various DDI strengths in component 1. The interaction parameters are $n_0 \tilde{g} = 2$ and $n_0 \tilde{g}_{12} = 1.8$.  Notice that as $n_0 \tilde{g}_d$ is increased, a short-range minimum emerges in $U_{(1,1)}(R)$.}
\end{figure}

We plot $U_{(1,1)}(R)$ for a range of $n_0 \tilde{g}_d$ with $n_0\tilde{g}=2$ and $n_0 \tilde{g}_{12} = 1.8 $ (corresponding to $\Delta = 0.19$) in Fig.~\ref{fig:hqvpot}. % For small $n_0 \tilde{g}_d$, the potential tends towards the $\propto \mathrm{ln}\frac{R}{\xi} / R^2$ form at large $R$, while 
For $n_0 \tilde{g}_d \gtrsim 1.2$, the potential flattens and begins to develop a minimum.  Thus, an attractive force emerges in the $(1,1)$ HQV system for a sufficiently strong DDI in component 1, implying that the HQV pairs % that emerge in the rotating binary condensate 
are indeed bound states of HQVs, or HQV \emph{molecules}.  

\begin{figure}[t]
\includegraphics[width=.95\columnwidth]{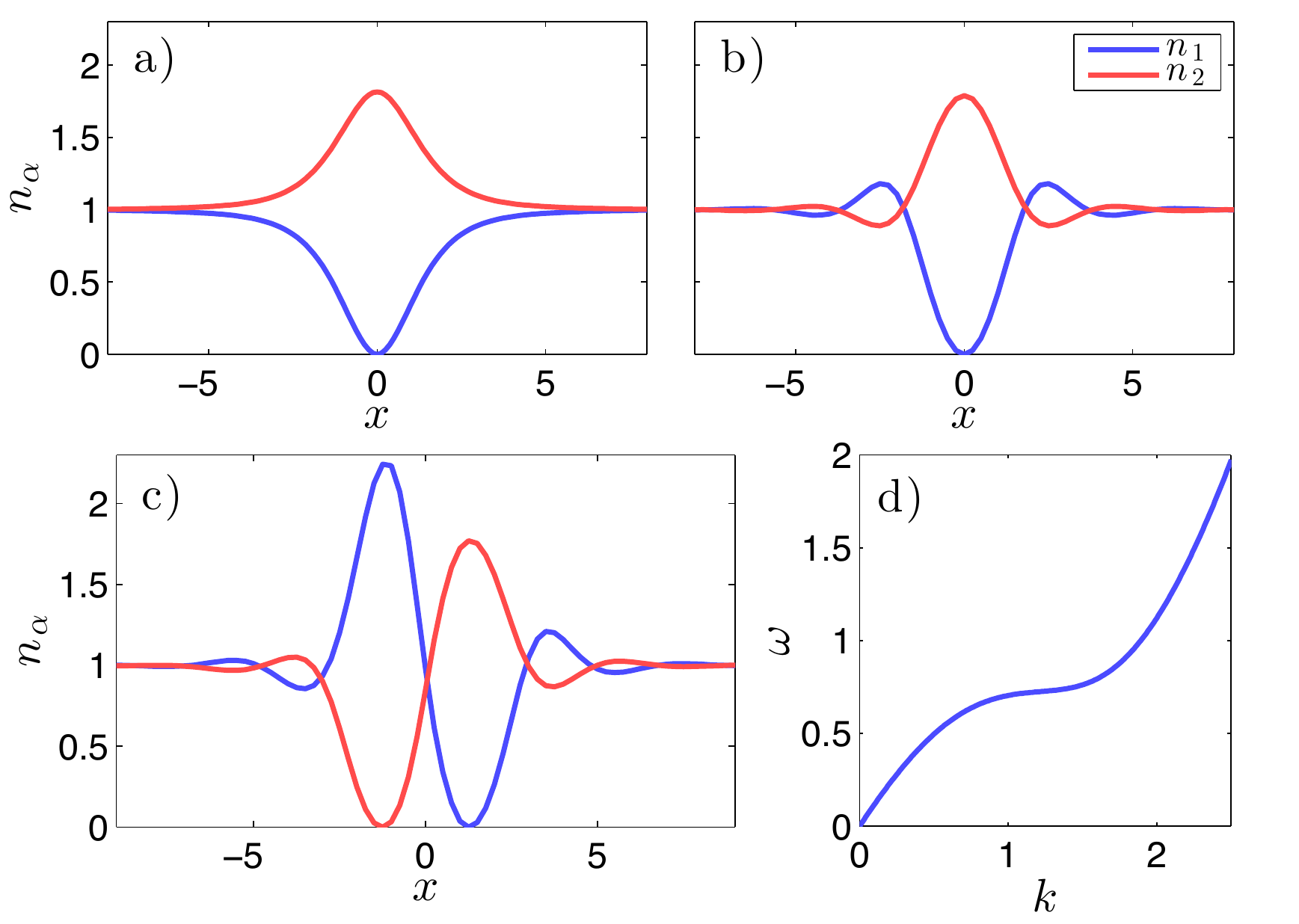}
\caption{\label{fig:prof} (color online).  (a)-(c) Density profiles $n_1(x,y=0)$ (blue lines) and $n_2(x,y=0)$ (red lines) for a homogeneous ($\lambda=\infty$) binary BEC with interaction parameters $n_0 \tilde{g} = 2$, $n_0 \tilde{g}_{12} = 1.8$, and (a) $n_0 \tilde{g}_d = 0$ (b)-(c) $n_0 \tilde{g}_d = 2.8$.  In (a)-(b), we show a $(1,0)$ HQV at $\rrho_{(1,0)}=0$.  Notice that the presence of dipolar interactions results in the emergence of a standing longitudinal spin wave near the HQV center.  In (c) we show a $(1,1)$ HQV system with $R=2.5$, corresponding to the minimum of $U_{(1,1)}(R)$ for these parameters.  (d) The spin-wave quasiparticle spectrum for the system in (b)-(c), but with no HQVs.
}
% (a)-(b) a $(1,0)$ HQV at $\rrho_{(1,0)}=0$ and (c) a $(1,1)$ HQV system with $R=2.5$, %$\rrho_{(1,0)} = 1.25 \mathbf{e}_x$ and $\rrho_{(0,1)} = -1.25 \mathbf{e}_{x}$, 
% corresponding to the minimum of $U_{(1,1)}(R)$ for these parameters.  Notice that the presence of dipolar interactions in (b) and (c) results in the presence of standing longitudinal spin waves near the HQVs.  d) The spin-wave quasiparticle spectrum for the system in (b)-(c), but with no HQVs.  }
\end{figure}

The emergence of this potential minimum can be understood intuitively by inspecting the density profiles of the HQVs at large $\tilde{g}_d$.  We plot the densities $n_\alpha(x,y=0)$ near a $(1,0)$ HQV located at $\rrho_{(1,0)} = 0$ for $n_0 \tilde{g}_{12} = 1.8$, $n_0 \tilde{g} = 2$, and $n_0 \tilde{g}_d = 2.8$ in Fig.~\ref{fig:prof}(b).  For comparison, we plot the densities for the analogous case with $\tilde{g}_d = 0$ in Fig.~\ref{fig:prof}(a).  In the dipolar case, components 1 and 2 exhibit out-of-phase radial density oscillations, or longitudinal spin oscillations on a length scale $\sim l_z$ near the HQV core.  A very similar feature emerges near a vortex in a single-component Q2D dipolar BEC~\cite{Yi06,Abad09,Wilson09}.  In this case, the density oscillations, or ``ripples,'' are due to the presence of a roton in the quasiparticle spectrum of the gas, which is coupled into the ground state density profile by the vortex core~\cite{Wilson08PRL}. Here, the situation is perfectly analogous: a spin-wave roton is coupled into the ground state of the binary BEC by the HQV core.  % Fig.~\ref{fig:prof}(b) demonstrates the analogous feature for a spin-wave roton near a HQV in a binary dipolar BEC.  
For reference, we plot the spin-wave branch of the quasiparticle spectrum (given by Eq.~(23) in Ref.~\cite{Wilson12}) for the spatially homogeneous binary condensate with these parameters % with $n_0\tilde{g} = n_0 \tilde{g}_d = 2$ and $n_0 \tilde{g}_{12} = 1.8$ 
in Fig.~\ref{fig:prof}(c).  Surprisingly, the rotonic feature in this spectrum is rather subtle, though the spin wave oscillations are quite distinct.  In Fig.~\ref{fig:prof}(d), we plot the densities for a $(1,1)$ HQV system separated by a distance $R\simeq 2.5 $, corresponding to the minimum of $U_{(1,1)}(R)$ for these parameters.  At this separation, the location of  the $(0,1)$ HQV core is commensurate with the peak in the $n_1$ density oscillation. %, suggesting that this configuration is energetically favorable.  % Thus, by forming at pair, the $(1,1)$ HQV system 
In other words, the formation of $(1,1)$ pairs allows the longitudinal spin to oscillate \emph{naturally} around the HQVs, resulting in an energetically favorable configuration.

Unlike the HQV molecules that emerge in coherently coupled binary BECs, the mutual attraction between the HQVs in this dipolar system does not depend on the relative phase of the components.  Thus, the $U_{(1,1)}(R)$ potential will also describe interactions in the $(-1,1)$ and $(1,-1)$ HQV systems, which correspond to HQV-antiHQV pairs, and are objects of fundamental importance in the BKT transition in multicomponent superfluids.

\emph{Discussion}--  Candidates for realizing the balanced system we discuss here include states in the ground hyperfine manifolds of atomic Cr, Er, or Dy, where components 1 and 2 could consist of states with spin projections $m_J = -J$ and $m_J = 0$, for example.  The $m_J = -J$ states of these bosonic atoms have magnetic dipole moments $6\,\mu_\mathrm{B}$, $7 \, \mu_\mathrm{B}$, and $10 \, \mu_\mathrm{B}$, respectively, where $\mu_\mathrm{B}$ is the Bohr magneton.  Though the relaxation lifetimes of states with $m_J>-J$ can be quite long~\cite{Griesmaier05}, such mixtures will be limited to lifetimes $\sim 10-100 \, \mathrm{ms}$ due to spin-exchange collisions, so experiments would have to operate on much shorter timescales.  However, we have verified that the HQV chain and molecule phases emerge for a range of mass and number imbalances as well, suggesting that these phases may be realized with appropriately engineered alkali-Cr or alkali-lanthanide mixtures, where the $\sim 1 \, \mu_\mathrm{B}$ magnetic dipole moment of the alkalis can be safely neglected.  For example, in a highly oblate trap with axial frequency $\omega_z = 2\pi \times 2 \, \mathrm{kHz}$, we find that HQV molecules emerge in a $^{87}$Rb-$^{164}$Dy mixture with integrated 2D densities $n_0 \sim 5\times 10^{10} \, \mathrm{cm}^{-2}$ near the trap center, assuming that the $s$-wave scattering lengths can be tuned by a Fano-Feshbach resonance~\cite{Chin10}.  

Finally, we have checked that our results hold when the condensate order parameters are not assumed to be separable by performing fully three-dimensional numerical calculations, indicating that the Q2D approximation provides a qualitatively accurate description of the HQV-pairing phenomena, even when the interaction energy scales are larger than $\hbar \omega_z$.  

%  for the fully three-dimensional case, where were drop the Q2D approximation and minimize $\langle \hat{\mathcal{H}} \rangle$ on a fully three-dimensional grid. 

% Though it is conceivable that the balanced case we study here could be realized using a mixture of $m_J=0$ and $m_J = \pm J$ hyperfine states of bosonic Cr, Er, or Dy, collisional spin-mixing would limit such mixtures to millisecond lifetimes. 

\emph{Conclusion}--  In summary, we studied the role of dipolar interactions in the ground state phase diagram of a rotating binary BEC in an oblate geometry.  This system exhibits an immiscible-miscible phase transition and a number of rich miscible phases that are characterized by the ordering of HQVs into various lattice configurations.  When one component is strongly dipolar and the other is non-dipolar, the interaction potential between $(1,0)$ and $(0,1)$ HQVs ($U_{(1,1)}(R)$) develops a minimum, resulting in an effective attraction between these HQVs at a short distance.  This leads to the emergence of a HQV chain phase and the formation of HQV molecules for sufficiently strong dipolar interactions.  In addition to playing a critical role in the ground state phase diagram of this system, the attractive HQV interactions should strongly influence its dynamical properties, including its superfluid turbulent behavior and BKT transition, which will be subjects of future work.

% This attractive HQV interaction should play a critical role not only in the ground state phase diagram, but also in the dynamics of this system, influencing, for example, the superfluid turbulent behavior and the BKT transition in the gas.  

%  If one component possesses a large dipole moment and the other is effectively non-dipolar, HQVs of opposite charge 

% For weak dipolar interactions, these lattices are either square (rhombic) or triangular in nature, resulting from strong HQV repulsion.  

% For strong dipolar interactions, however, the $U_{(1,1)}(R)$ interaction potential flattens and eventually develops a minimum, leading to the effective attraction of $(1,0)$ and $(0,1)$ HQVs at short distances.  

% In oblate geometries and for sufficiently strong dipolar interactions, 

\emph{Acknowledgments}--  We thank Ashton Bradley and Tom Billam for sharing their vortex identification algorithm.  RMW acknowledges support from an NRC postdoctoral fellowship.  This material is based upon work supported by the NSF under Grant No. PHY-1004975, and was partially supported by the NSF under the Physics Frontiers Center Grant No. PHY-0822671 and the ARO Atomtronics MURI.

%\begin{figure}[h]
%\includegraphics[width=.8\columnwidth]{tiltdensml.pdf}
%\caption{\label{fig:phases} (color online).  Example figure  }
%\end{figure}
%
%\begin{figure}[h]
%\includegraphics[width=.8\columnwidth]{hqvpottilt.pdf}
%\caption{\label{fig:phases} (color online).  Example figure  }
%\end{figure}

% \bibliography{hqv}

\end{document}